\documentclass[aps,pre,preprint,superscriptaddress]{revtex4-1}

\usepackage{amssymb}
\usepackage[dvips]{graphicx} 
\usepackage[usenames, dvipsnames]{color}
\usepackage[version=3]{mhchem}
\usepackage{amsmath}
\usepackage{wasysym}

\begin{document}


\title{Acid induced assembly of a reconstituted silk protein system}


\author{A. Pasha Tabatabai}
\affiliation{Georgetown University Department of Physics and Institute
  for Soft Matter Synthesis and Metrology, Washington DC 20057}
\author{Katie M. Weigandt}
\affiliation{Center for Neutron Research, National Institute of Standards and Technology, Gaithersburg MD 20899}
\author{Daniel L. Blair}
\affiliation{Georgetown University Department of Physics and Institute
  for Soft Matter Synthesis and Metrology, Washington DC 20057}
\email{daniel.blair@georgetown.edu}


\date{\today}

\begin{abstract}
Silk cocoons are reconstituted into an aqueous suspension, and protein stability is investigated by comparing the protein's response to hydrochloric acid and sodium chloride. Aggregation occurs at $<8$ mM hydrochloric acid that is not correlated to protein protonation, while sodium chloride over the same range of concentrations does not cause aggregation.
We measure the structures present on the protein and aggregate lengthscales in these solutions using both optical and neutron scattering, while
mass spectrometry techniques shed light on a possible mechanism for aggregate formation.
We find that the introduction of acid modulates the aggregate size and pervaded volume of the protein, an effect that is not observed with salt.
\end{abstract}


\maketitle

\section{Introduction}
Silk protein materials are the focus of many current biomedical applications because they are strong and biocompatible~\cite{Altman:2003ee, Wang:2006tn}. Additionally, silk materials can now be made into non-fibrous forms by taking advantage of protein reconstitution techniques; pre-spun cocoon fibers are chemically dissolved into solution in large quantities~\cite{Rockwood:2011gb}. However, the protein network morphologies of new silk materials are different than that of the pre-spun fiber. Consequently, the optimization of new silk materials requires an understanding of the fundamental interactions between constituent particles.

One novel material is a gel formed by running a DC electric current through reconstituted silk; the acidic domain created during electrolysis prompts the assembly of protein into a soft gel~\cite{Leisk:2010fx, Yucel:2010jl}. Electric field generated gels exhibit interesting bulk mechanical features that are a consequence of interactions between individual protein~\cite{Tabatabai:2014id}. In order to understand this gelation phenomenon better, we controllably add hydrochloric acid (HCl) at concentrations $c_H$ to reconstituted silk solutions. Protein aggregates form at $c_H<8$ mM (M=mol L$^{-1}$).
Protein networks form for $c_H>8$ mM but phase separate; quantitative analysis is not done in this regime. In many charged particle systems, pH changes induce aggregation through protonation of the particles, minimizing electrostatic repulsion. However, we will show that aggregation is not caused by protonation of amino acid side-chains.

Sodium chloride (NaCl) is added to the protein at concentrations $c_N$ in the same range as $c_H$.
A comparison between HCl and NaCl effects reveals that HCl does not induce aggregation through charge screening.

In this work, we focus on the structural changes and interactions of reconstituted silk fibroin protein in the presence of NaCl or HCl to understand the mechanism by which silk proteins assemble. 
This mechanism is investigated by determining the structure of the protein through measurement of the fractal dimension. Sub-protein lengthscales are not easily accessible by most measurement techniques but can be resolved using SANS~\cite{Weigandt:2009je, Weigandt:2011ji}. An understanding of the protein structure at both the aggregate and sub-protein lengthscales provides insight on the fundamental interactions of reconstituted silk and is an essential step in correlating microstructural interactions with bulk mechanical properties.


%
%

\section{Methods and materials}
\subsection{Reconstituted protein}
Silk protein solutions are reconstituted from \textit{Bombyx mori} silk cocoons following the method detailed by Rockwood et al.~\cite{Rockwood:2011gb}. Cocoons are boiled in aqueous sodium carbonate for 10 minutes to remove the globular sericin, leaving behind only the insoluble structural fibroin proteins that are washed in deionized water. 

The fibers are soaked in an aqueous solution of lithium bromide (LiBr) at 70 $^{\circ}$C for 2 hours to solubilize the protein
through denaturation. The LiBr-protein solution is dialyzed against cycled deionized water for 48 hours in a dialysis bag with a molecular weight cutoff of 10 kDa to remove the solubilized ions. Undissolved fibers are removed by centrifugation, and protein aggregates larger than 0.45 $\mu$m are removed through filtration. The resultant reconstituted silk solution is composed entirely of the fibroin protein and will now be referred to as silk protein.

Silk is stable in an unbuffered solution for weeks/months providing a stable high concentration ($\sim 50$ mg\ mL$^{-1}$) feedstock. Silk proteins are monomerized during the reconstitution process and have a well characterized polydispersity~\cite{Partlow:2016va}.
The amino acid sequence of the protein is primarily low complexity repeats of small side chain amino acids; the protein can be approximated as a polymer. 
The final silk protein concentration is determined using ultraviolet-visibile light (UV-Vis) spectroscopy; the molar extinction coefficient at 280 nm is 441030 cm$^{-1}$M$^{-1}$ as determined from the amino acid sequence~\cite{Zhou:2001gk}.
For neutron scattering measurements, aqueous silk solutions are subsequently dialyzed against deuterium oxide (D$_2$O) to an H$_2$O:D$_2$O solvent volume ratio of $5:95$.

\subsection{Mixing silk with HCl or NaCl}
NaCl and HCl are chosen because they are common, ionize completely at our experimental concentrations, and share a common anion.
HCl and NaCl solutions are prepared by dilution in either H$_2$O or D$_2$O.
For neutron scattering, HCl stock solutions are prepared by dilution of a 37\% HCl in H$_2$O assay into D$_2$O. At $c_H< 5$ mM, solutions of HCl in D$_2$O have a scattering length density equivalent to 100\% D$_2$O;
further dilutions of this stock HCl solution do not appreciably change the scattering length density. Acid or salt solution is added to the D$_2$O protein solution at a 1:1 volume ratio resulting in an H$_2$O:D$_2$O solvent volume ratio of $2.5:97.5$.

The exact stoichiometric interaction between HCl or NaCl and silk protein is unknown so we define the molar equivalent (ME) as the ratio of the moles of added compound to the moles of silk in solution. 
Solutions are prepared in deionized water with background NaCl present in trace amounts and at pH 9 ($\mathrm{[HCl]} = 10^{-9}$ M). Since $c_H, c_N\geq0.1$ mM and are much larger than their respective background concentrations, ME is equivalent to the absolute ion concentration per protein.

\subsection{Measuring ion concentrations}
The concentration of lithium (Li) and bromine (Br) ions are measured with inductively coupled plasma mass spectrometry (ICP-MS) for the isotopes Li-6, Li-7, Br-79, and Br-81; reported values are a sum of these isotopes. The count rates from an ICP-MS measurement are converted into a concentration by measurements of known LiBr concentrations; the conversion between count rates and concentration is confirmed to be linear in the measured range and least squares fitting gives the conversion factor with 95\% confidence.
Data plotted are the best estimates, and error bars correspond to the propagation of uncertainties in the conversion factors. 

Using centrifugation filters with a molecular weight cutoff of 10 kDa, elutions of silk solutions containing HCl or NaCl are collected. The protein and ion concentrations in the elutions are measured with UV-Vis spectroscopy and ICP-MS respectively. Changes in elution ion concentrations are representative of changes in mobile ion concentrations in the bulk.

\section{Results and discussion}

\subsection{Aggregate formation}

\begin{figure}[]
\centering
  \includegraphics[width=0.5\linewidth]{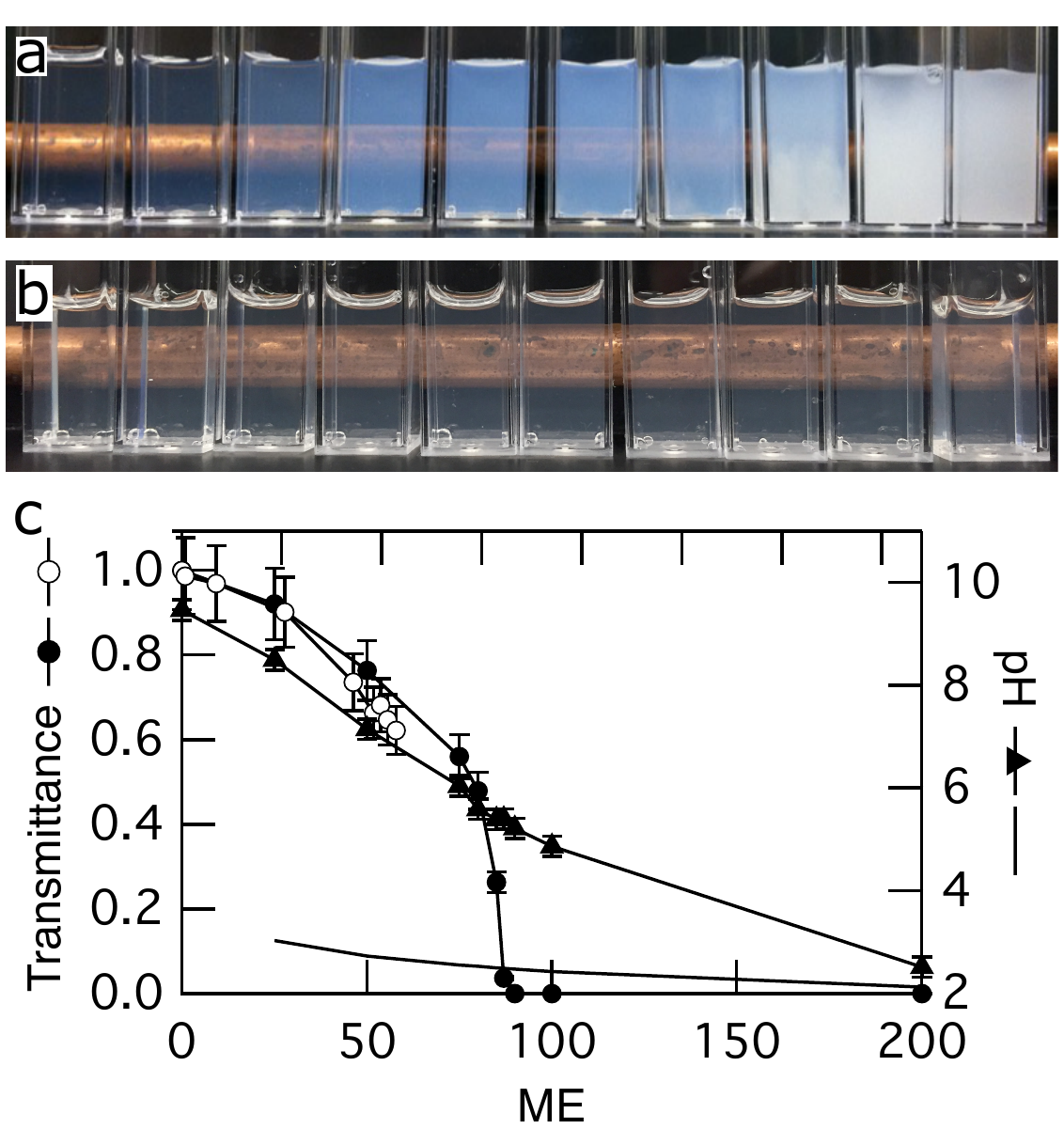}
  \caption{ $37\ \mu$M silk protein solutions with (a) $c_H= 0\to7.4$ mM (left to right) ($0\le \mathrm{ME}\le200$) and (b) NaCl over the same range $0\le \mathrm{ME}\le200$. (c) Data corresponding to samples from (a): the normalized transmittance for each sample at 488 nm ($\CIRCLE$), the measured pH ($\blacktriangle$), and the expected pH value ($-$). Transmittance for the samples used in the neutron scattering experiment ($\Circle$).}
  \label{fgr:turbidity}
\end{figure}

The turbidity of silk solutions increases as $c_H$ increases from $0\to7.4$~mM [Fig. \ref{fgr:turbidity}(a)]. The protein concentration $c_P=37\ \mu$M, giving $0\leq\mathrm{ME}\leq 200$. Transmittance $T$ decreases rapidly from $T = 60\%$ to $T=0\%$ in the range of $75\le\mathrm{ME}\le90$ and remains at zero with any additional acid [Fig. \ref{fgr:turbidity}(c)]. In contrast, NaCl has no visible effect on the sample turbidity at $0\leq\mathrm{ME}\leq 200$ [Fig.~\ref{fgr:turbidity}(b)]. The difference in total optical scattering shows that HCl and NaCl affect reconstituted silk differently. Inter-protein aggregation is a consequence of HCl induced association.

The pH of each of the samples in Fig.~\ref{fgr:turbidity}(a) are measured directly with a pH probe and decrease smoothly from $9\to2$ with increasing ME [Fig.~\ref{fgr:turbidity}(c)].
Since $\sim1$~\% of the amino acids on the protein are ionizable, the approximation for the pH of the solution $\mathrm{pH}=-\mathrm{log}(c_H)$ predicts a decrease in pH from $3 \to 2$. However, the measured pH differs significantly from the predicted value due to
the strong association of H$^+$ with the protein.
When $\mathrm{ME} =200$, the measured pH and the expected pH value begin to converge; the protein is saturated with H$^+$ and unable to continue to buffer the solution. Consequently, any additional H$^+$ ($\mathrm{ME}>200$) will remain in solution. 
The pH value associated with the dramatic change in turbidity is not correlated with
any amino acid side chain pKa nor does the pH change stepwise as ionization occurs. 
Additionally, the sample is completely turbid at pH values greater than the silk isoelectric point 
(values in the literature range from $3.2\to4.2$). Therefore, HCl induced aggregation is not due to a minimization of charge repulsion~\cite{Cadwallader:1946do, Kim:2004fp, Terry:2004vh, Lin:2013hi}.

\begin{figure}[h]
\centering
  \includegraphics[width=0.5\linewidth]{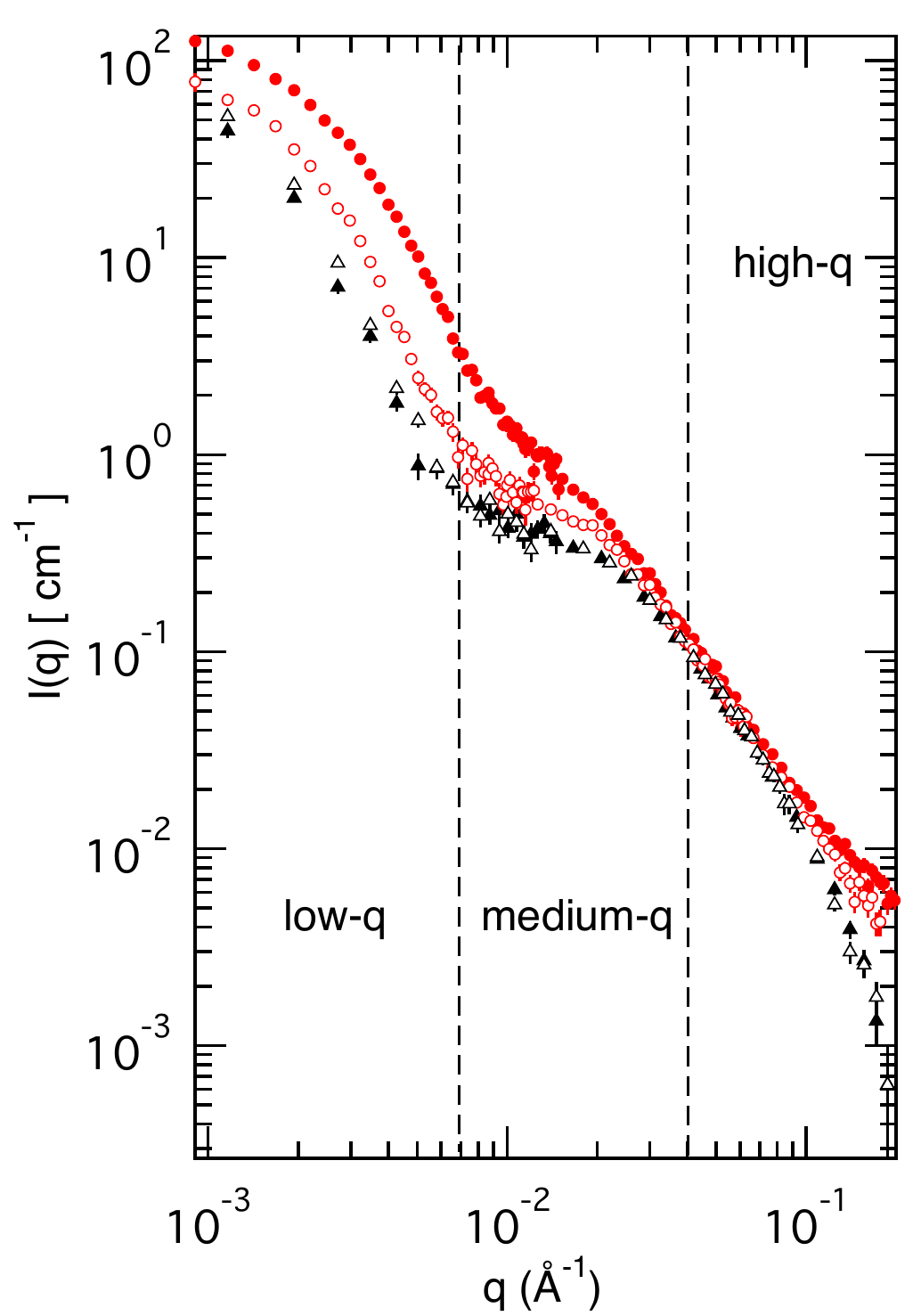}
  \caption{Scattering intensity $I(q)$ of deuterated silk samples at $c_P=37\ \mu$M: 0.5 mM NaCl ($\triangle$), 1.0 mM NaCl ($\blacktriangle$), 0.5 mM HCl (\textcolor{red}{$\Circle$}), and 1.0 mM HCl (\textcolor{red}{$\CIRCLE$}). }
  \label{fgr:sva}
\end{figure}

Changes in $c_H$ result in changes in $T$. Likewise, changes in $c_H$ alter the intensities of SANS spectra on both the aggregate and protein lengthscales [Fig.~\ref{fgr:sva}]. An increase in $c_H$ 
results in an increase in the scattering intensity $I$ for wavevector $q$ in the low-$q$ ($q<7\times10^{-3}$\AA$^{-1}$) regime.
The low-$q$ increase signifies an increase in the total aggregate volume that is consistent with turbidity measurements [Fig.~\ref{fgr:turbidity}]. High-$q$ ($q>4\times10^{-2}$\AA$^{-1}$) scattering is independent of $c_H$, while a slope change between high-$q$ and medium-$q$ ($7\times10^{-3}$\AA$^{-1}<q<4\times10^{-2}$\AA$^{-1}$) provides the characteristic lengthscale for a single protein.

In contrast, $I(q)$ are independent of changes in $c_N$; unchanged low-$q$ scattering is consistent with the lack of turbidity in Fig.~\ref{fgr:turbidity}(b). Instead of inducing aggregation, charge screening effects from NaCl appear at $q>0.06$~\AA$^{-1}$ as observed by a slight increase in the slope of the NaCl curves in Fig.~\ref{fgr:sva}.
The presence of $q$ dependent scattering at low-$q$ for these spectra is a consequence of unavoidable aggregate byproducts from reconstitution. However, \textit{changes} in aggregates are only induced by HCl.

Because $c_H$ and $c_N$ span the same range and aggregates form only from increases in $c_H$, aggregation is only induced by HCl. Aggregation is not a consequence of ionic strength or charge screening; incorporation of HCl is needed before inter-protein associations can exist.

\subsection{Reconstituted protein stability} 

In spun fibers, silk is folded into multiple $\beta$-sheets through hydrogen bonding: a dipole-dipole interaction between polar amine \ce{N\bond{-}H} and carbonyl \ce{C\bond{=}O} groups on the protein backbone.
However, reconstituted silk is predominantly an unstructured random coil where \ce{N\bond{-}H} and \ce{C\bond{=}O} are unassociated~\cite{Kim:2004fp, Partlow:2016va}.
We find that the random coil configuration is stable because of residual lithium ions from the reconstitution process that associated with \ce{C\bond{=}O} analogous to a charge-dipole interaction. A comparison of lithium concentrations before and after HCl or NaCl addition indicates a mechanism for aggregate formation.

\begin{figure}[h]
\centering
  \includegraphics[width=0.5\linewidth]{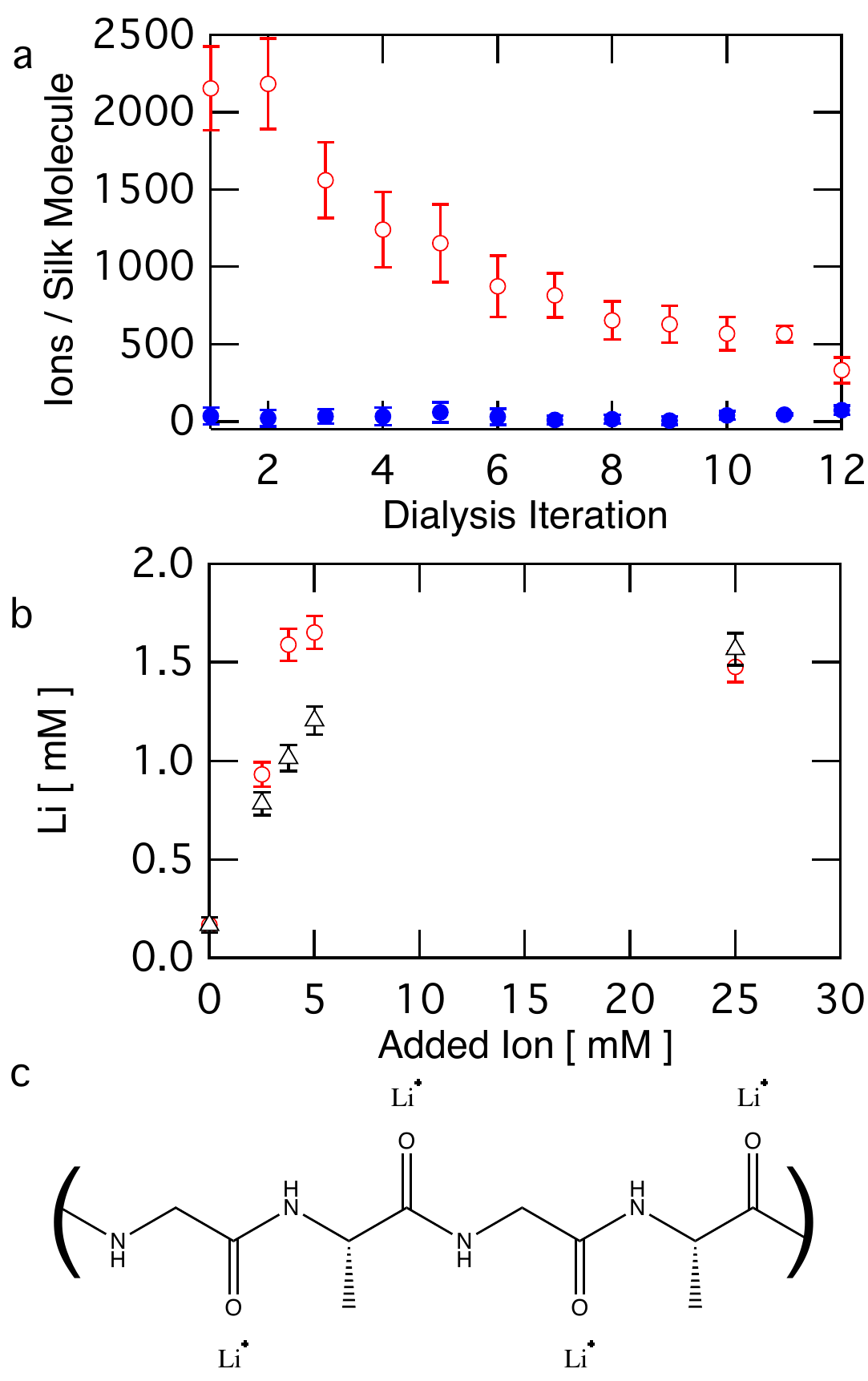}
  \caption{(a) Lithium (\textcolor{red}{$\Circle$}) and bromine (\textcolor{blue}{$\CIRCLE$}) ions per protein during twelve extra dialysis iterations beyond the normal reconstitution protocol. Iteration number one represents the ion concentrations present after the standard reconstitution protocol. (b) Lithium ions in the elution with the addition of either HCl (\textcolor{red}{$\Circle$}) or NaCl ($\bigtriangleup$). (c) A common segment of the protein sequence (G-A-G-A) with lithium ions drawn to associate with carbonyl oxygens.}
  \label{fgr:lithium}
\end{figure}

The number of Li$^+$ per protein after the reconstitution protocol is $2200\pm300$ [Fig.~\ref{fgr:lithium}(a)]. The silk protein has 5263 amino acids, therefore there is one Li$^+$ for every $2.4\pm0.3$ amino acids, which is equivalent to one Li$^+$ for every $2.4\pm0.3$ carbonyl oxygens on the protein backbone. This ratio is consistent with molecular dynamics simulations and density functional theory calculations for Li$^+$ associations with electronegative oxygens on different polymer chemistries~\cite{Webb:2015vf, Russo:2001vt, Jover:2008jq}.
Therefore, Li$^+$ are present at the right stoichiometric ratio to be associated to the carbonyl oxygens on the protein backbone. 
Li$^+$ has previously been found through experiments and computation to have a high binding affinity with the carbonyl oxygen on the protein mimetic molecule N-methyl-acetamide as well as with the amino acids~\cite{Jover:2008jq, Manin:2016el}. Associations of Li$^+$ to the carbonyl is reasonable since chaotropic LiBr is chosen in the reconstitution process to denature the protein, i.e., disassociate \ce{N\bond{-}H} and \ce{C\bond{=}O} groups. 
Silk's stability/inability to self-assemble is likely caused by the continued complexation of the carbonyl with Li$^+$~\cite{Koebley:2015il}.
Meanwhile, the Br$^-$ concentration is very low; presumably all Br$^-$ are removed during dialysis.

To confirm that Li$^+$ are associated to the protein, the concentration of of Li$^+$ per protein is measured during additional dialysis iterations. The silk solution $\mathrm{V}_{\mathrm{silk}}=15$~mL is placed back into a 10 kDa dialysis cassette and set in a reservoir $\mathrm{V}_{\mathrm{res}}=3000$~mL of gently stirred deionized water. After $24$ hours, an aliquot of the silk solution is collected, and the reservoir water is replaced with new deionized water. After 12 iterations, the number of Li$^+$ per protein decreases approximately by a factor of 10, whereas Br$^-$ concentrations remain at zero [Fig.~\ref{fgr:lithium}(a)].

The concentration of Li$^+$ inside the cassette (c$_1$) is greater than the concentration in the reservoir initially.
After $24$ hours, the new equilibrium concentration of Li$^+$ in the cassette 
$\mathrm{c}_2 = \mathrm{c}_1\times \mathrm{V}_{\mathrm{silk}}/ \mathrm{V}_{\mathrm{res}}$.
Following twelve dialysis iterations, the recursive relation gives 
$\mathrm{c}_{12}=\mathrm{c}_1\times \left(\mathrm{V}_{\mathrm{silk}}/ \mathrm{V}_{\mathrm{res}}\right)^{11}$, or $\mathrm{c}_{12}/\mathrm{c}_1\sim \left(1/200\right)^{11}$. 
The expected concentration ratio $\mathrm{c}_{12}/\mathrm{c}_{1}$ is significantly different than the measured ratio $\mathrm{c}_{12}/\mathrm{c}_{1}\sim 1/10$. Therefore, it must be the case that Li$^+$ has an affinity to the silk protein.

We vary $0\ \mathrm{mM} \leq c_H,c_N \leq 25\ \mathrm{mM}$
in silk solutions prepared with the standard reconstitution protocol to determine if H$^+$ and Na$^+$ ions displace Li$^+$ from the protein.
 As $c_H$ of $c_N$ increases, the concentration of free Li$^+$ increases [Fig.~\ref{fgr:lithium}(b)].
HCl is better at displacing Li$^+$ than NaCl at low concentrations, but there is no effective difference at 25 mM. 
Because both HCl and NaCl remove Li$^+$ from the protein, we know that both compounds interact with the protein. H$^+$ associated to the protein is a likely mechanism for the discrepancy between the measured pH value and the expected value shown in Fig.~\ref{fgr:turbidity}.

Since Li$^+$ is known to associate with carbonyls, as previously mentioned, it is likely that both H$^+$ and Na$^+$ also associate with the carbonyls to conserve charge.
The displacement of Li$^+$ with either H$^+$ or Na$^+$ is reasonable given that density functional theory calculations provide roughly equivalent affinities of H$^+$, Na$^+$, and Li$^+$ for carbonyl oxygens~\cite{Jover:2008jq, Gorman:1992wy, Wyttenbach:2000wq}. It is not only the removal of Li$^+$ that causes aggregation but specifically the removal of Li$^+$ by H$^+$ that is needed to destabilize the protein an induce aggregation.

\subsection{Acid induced structures}
Our results establish a link between aggregate formation and acid. HCl induced protein aggregates are now measured in order quantify HCl dependent changes in structure. Silk protein solutions at a final concentration of $c_P=54\ \mu$M protein in D$_2$O are mixed with HCl in the range of $0\ \mathrm{mM}\le c_H\le3.125\ \mathrm{mM}$, giving $0\le \mathrm{ME}\le 58$. The transmittance of these samples is a function of ME and ranges from $100\%\to50\%$ [Fig.~\ref{fgr:turbidity}].

\begin{figure}[h]
\centering
  \includegraphics[width=0.5\linewidth]{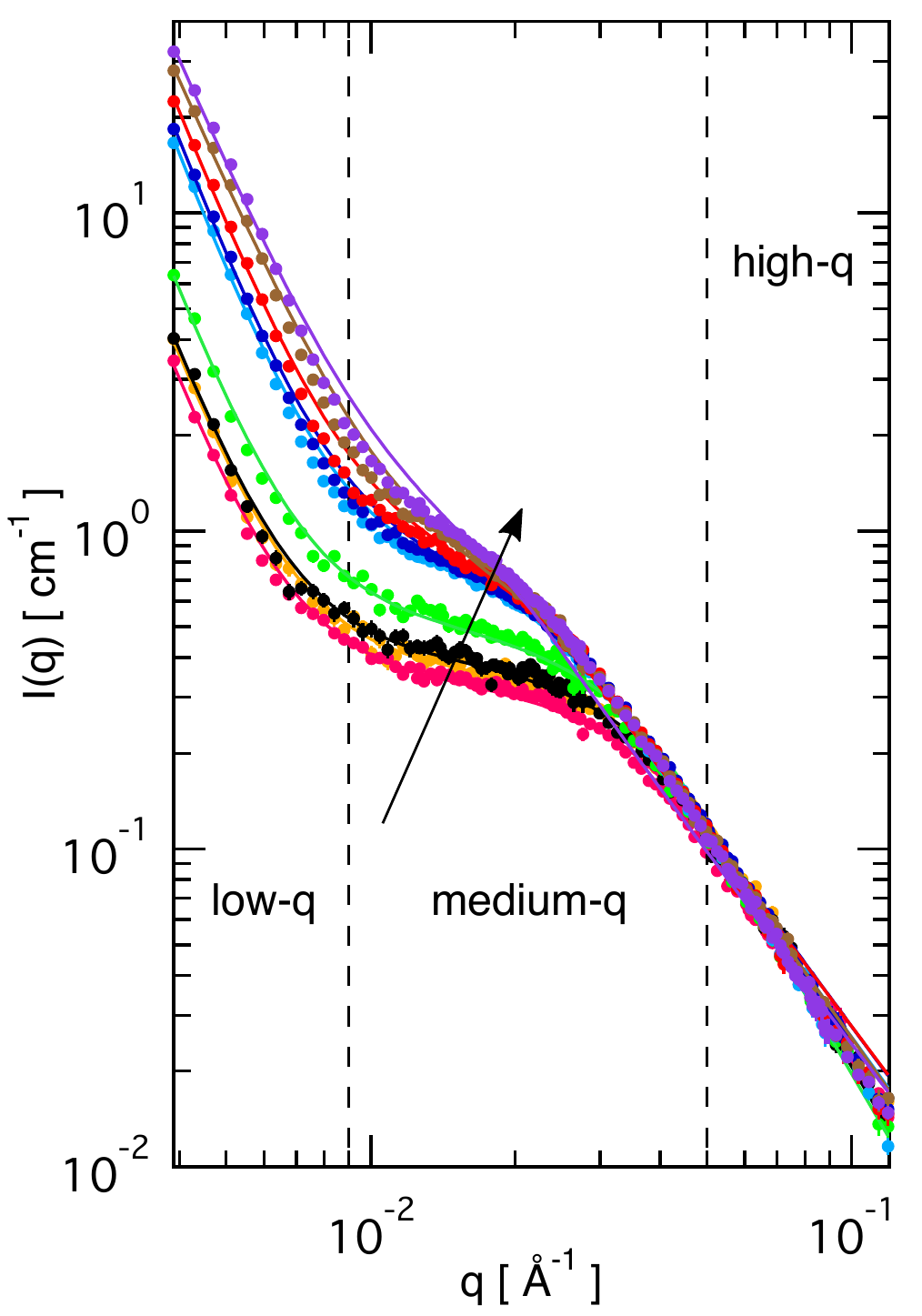}
  \caption{$I(q)$ changes monotonically as ME increases along the arrow from $0\to58$ except for the two lowest values of ME: $\mathrm{ME}=0$ (\textcolor{YellowOrange}{$\CIRCLE$}) and $\mathrm{ME}=1.9$ (\textcolor{Magenta}{$\CIRCLE$}). $I(q)$ for each value of ME are fit with a double Guinier Porod model and are plotted as the solid lines.}
  \label{fgr:Iq}
\end{figure}

Changes in the scattering intensity $I(q)$ reveal ME dependent structural changes [Fig.~\ref{fgr:Iq}]. 
At high-$q$ ($q>4\times10^{-2}$\AA$^{-1}$), $I(q)$ is unchanged over the range of ME; the protein has a constant fractal dimension irrespective of the aggregate conditions.
Medium-$q$ scattering ($9\times10^{-3}$\AA$^{-1}$$<q<4\times10^{-2}$\AA$^{-1}$) exhibits an ME dependent slope change that likely exists due to the superposition of scattering from the aggregate and protein lengthscales. 
Although byproducts of reconstitution are seen in the $q$ dependence at low-$q$  ($q<9\times10^{-3}$\AA$^{-1}$) for $\mathrm{ME}=0$, increases in ME manifest themselves as a vertical shift in $I(q)$; more aggregates are formed with increasing ME as seen in Fig.~\ref{fgr:turbidity}. ME does not affect the low-$q$ slope; the internal structure of an aggregate remains constant.

$I(q)$ in Fig.~\ref{fgr:Iq} are a result of scattering from both the individual proteins and aggregates. To minimize assumptions during data fitting, each lengthscale is interpreted using the Guinier Porod model in three dimensions,
\begin{equation}
I(q)=
\begin{cases}
G\cdot\exp({\frac{-q^2R_G^2}{3}}),& \text{if } q\leq q_1\\
\frac{D}{q^d},& \text{if } q\geq q_1
\end{cases}
\label{me:GP}
\end{equation}
where G is the Guinier scaling parameter, D the Porod scaling factor, $d$ the Porod exponent, and characteristic length $q_1$~\cite{Hammouda:2010em}.
 Both of these equations and their derivatives must be continuous at $q=q_1$, therefore the following relations must hold:
\begin{equation}
\begin{split}
& q_1=\frac{1}{R_G}\left(\frac{3d}{2}\right)^{\frac{1}{2}},\\
& D=G\exp{\left(-\frac{d}{2}\right)}\left(\frac{3d}{2}\right)^{\frac{d}{2}}\frac{1}{R_G^d}.
\end{split}
\end{equation}
The Guinier regime of the protein and the Porod regime of the aggregate are not well separated, so both the aggregate and the protein are fit simultaneously using the same functional form, giving
\begin{equation}
I(q)_\mathrm{total}=I(q)_\mathrm{protein}+I(q)_\mathrm{aggregate}.
\label{me:doubleGP}
\end{equation}

\begin{figure}[h]
\centering
  \includegraphics[width=0.5\linewidth]{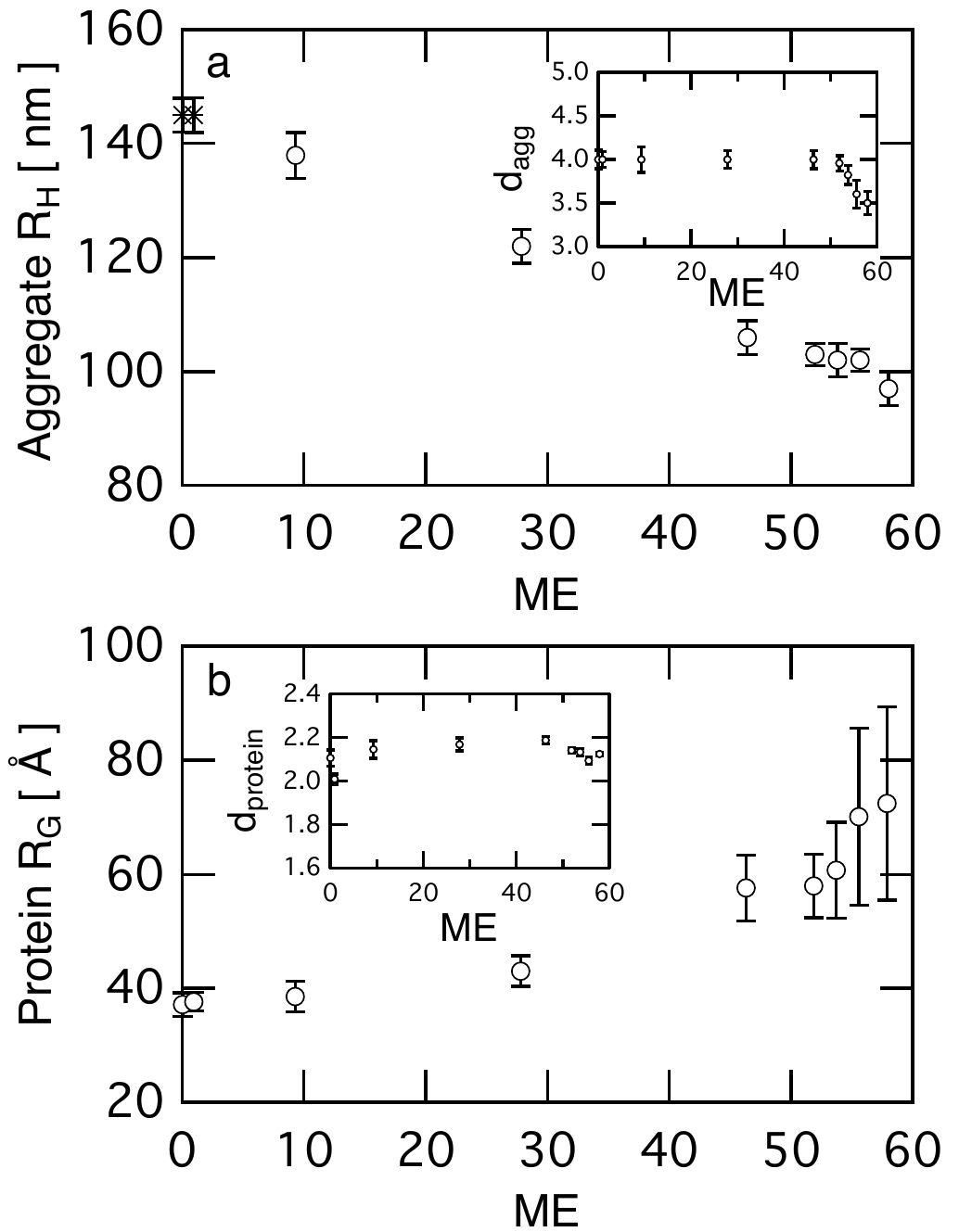}
  \caption{(a) Aggregate $R_H$ as determined from DLS as a function of ME. The measured samples for $\mathrm{ME}>9$ ($\Circle$) are used to approximate the hydrodynamic radii of the two unmeasured samples ($\ast$) by linear interpolation, and the error bars correspond to the propagation of uncertainty in the interpolation. (a-inset) $d_\mathrm{aggregate}$ determined by the double Guinier Porod fit of I$(q)$. (b) Protein $R_G$ as determined through the double Guinier Porod fit. (b-inset) $d_\mathrm{protein}$ determined by the single Guinier Porod fit.}
  \label{fgr:DLSfit}
\end{figure}

This double Guinier Porod fit has six free variables: $R_G$, G, and $d$ for both the protein and the aggregate. To minimize the number of free fitting variables, DLS is used to identify the hydrodynamic radius $R_H$ of the aggregates, which is used as the aggregate $R_G$. The aggregate radii decrease from $145\to 95$ nm with increasing ME [Fig.~\ref{fgr:DLSfit}(a)]. Since the increase in turbidity coincides with a decrease in aggregate size, the number of aggregates must increase with ME. It is likely that the original $\mathrm{ME}=0$ aggregates persist, and the number of smaller aggregates increases with ME.

Additionally, $d_\mathrm{protein}$ is determined by fitting only over the protein lengthscale ($q>1.5~\times10^{-2}$~\AA$^{-1}$) using Equation~\ref{me:GP} and is independent of ME [Fig.~\ref{fgr:DLSfit}].
Taking the distributions of best estimates, we find $d_\mathrm{protein}=2.12\pm0.02$; over this range of ME, the fractal dimension of the protein is equivalent to a polymer on the bad side of $\theta$-solvent conditions.

Each scattering curve in Fig.~\ref{fgr:Iq} is fit using Equations~\ref{me:GP}-\ref{me:doubleGP} and the fitting results are plotted as solid lines. Double Guinier Porod fitting provides the best estimates and uncertainty for the Porod exponent of the aggregate $d_\mathrm{aggregate}$ and $R_G$ of the protein. As ME increases, $R_G$ for the protein increases from $40\to70$ \AA \ [Fig.~\ref{fgr:DLSfit}(b)]. The value of $R_G$ for $\mathrm{ME}=0$ is consistent with the only other known neutron study of reconstituted silk~\cite{Greving:2010jc}.
$d_\mathrm{aggregate}\sim4$ for most of the values of ME, but varies slightly at high values of ME when the fitting quality in Fig.~\ref{fgr:Iq} starts to decline [Fig.~\ref{fgr:DLSfit}(a)]; the aggregates amongst all values of ME are similarly structured and dense.

The measured values of $R_H$ are close to the experimentally accessible lengthscale of SANS. The beginning of a turnover at low-$q$ is seen in Fig.~\ref{fgr:sva} for samples with an extended $q$ range, as expected from the DLS data, but the complete low-$q$ plateau is still out of the range of SANS. The low-$q$ plateau is accessible with ultra-small angle neutron scattering (USANS).
However, combining SANS and DLS minimizes stability/aging effects of the protein that could be observed during a long USANS experiment.

\section{Conclusions}
Silk protein in reconstituted solutions are stabilized by Li$^+$ associated to a fraction of carbonyl oxygens.
The addition of either HCl or NaCl displaces bound Li$^+$, but multi-protein structures are capable of forming only in the presence of HCl. 
The removal of Li$^+$ from the \ce{C\bond{=}O} dipole suggests that aggregation involves the reassociation of \ce{C\bond{=}O} and \ce{N\bond{-}H} as is seen in the silk fiber~\cite{BouletAudet:2011gu}. However, the replacement of Li$^+$ with H$^+$ leads us to suggest an alternative bond:
H$^+$ facilitates a bond between two separate \ce{C\bond{=}O} for an effective dipole-charge-dipole interaction. This complex is called a bifurcated hydrogen bond and is observed in biological systems~\cite{Feldblum:2014tq}. It will be important to distinguish between hydrogen and bifurcated hydrogen bonding because the two types of interactions have different binding strengths~\cite{Feldblum:2014tq}.

While bonds cannot be seen explicitly, changes in the structure of the protein are indirect evidence of changing protein-protein interactions. We show that ME increases the protein $R_G$ in an aggregate but not the protein structure defined by $d_\mathrm{protein}$. At first glance, these results appear to violate mass conservation, since the size of a particle can not increase without changing its density.
However, these results are possible if either 1) the boundary between one protein and its aggregated partner is undefined because of intercalation, or 2) the proteins are extending. Because $d_\mathrm{aggregate}\sim4$, the most likely scenario is that two proteins intercalate, allowing the formation of a dense aggregate. Importantly, we have shown that reconstituted proteins do not form ordered structures in an aggregate, and that aggregates exhibit a characteristic size that depends on ME. These results provide an important link in understanding reconstituted silk aggregation and a first step in correlating silk microstructure with mechanical properties.

%
%

\section{Acknowledgements}
This work was supported by the Air Force Office of Scientific Research through grant no. FA9550-07-1-0130. APT would also like to thank the Walter G. Mayer Endowed Scholarship Fund for the support.
This work utilized facilities supported in part by the National Science Foundation under Agreement No. DMR-1508249.
This work benefited from the use of the SasView application, originally
developed under NSF award DMR-0520547. SasView contains code developed with
funding from the European Union's Horizon 2020 research and innovation program
under the SINE2020 project, grant agreement No 654000.

\appendix
\section{Small angle neutron scattering}

Neutron scattering experiments are performed on the 30 m SANS instruments at the National Institute of Standards and Technology Center for Neutron Research~\cite{Glinka:1998wy, Mildner:2005ts, Cheng:2000wo, Cook:2005jx, Barker:2005tz}.
Measurements of the scattering intensity $I$ as a function of wavevector $q$ for the range $10^{-3}$~\AA$^{-1}$~$<q<0.5$~\AA$^{-1}$
are accomplished by using both 6 \AA\ wavelength neutrons at detector positions of 1 m, 4 m, and 13 m and 8.09 \AA\ lens focused neutrons at a detector position of 15.3 m. Deuterated silk solutions are pipetted into rectangular quartz cuvettes and mixed rapidly with an equal volume of HCl or NaCl solution. To ensure we measure the steady-state structure, samples equilibrate for 12 hours before the measurement at ambient conditions.

Data are reduced using the Igor macros, and the $q$-independent background scattering intensity for each sample is averaged and subtracted from each curve before plotting so that data represent scattering from the protein only~\cite{Kline:2006fh}.
Curve fitting is performed using SasView and custom code where a nonlinear least squares regression is fit iteratively for the best estimate of each fitting parameter as well as the error associated with 95\% confidence for each value~\cite{SasViewversion:we}.

\section{Dynamic light scattering}
The hydrodynamic radius $R_H$ of an aggregate is determined by
dynamic light scattering (DLS) on the same silk solutions from SANS over the range $5\times10^{-4}$~\AA$^{-1}$~$<q<25\times10^{-4}$~\AA$^{-1}$ and 
has a minimum at $q=25\times10^{-4}$~\AA$^{-1}$. Reported values of $R_H$ are calculated at $q=25\times10^{-4}$~\AA$^{-1}$, and a range of $R_H$ is determined through five measurements of the same sample.

\section{Transmittance}
Transmittance values are calculated by measuring the absorbance $A$ of the silk solution at 488 nm as measured by UV-Vis spectroscopy. The transmittance
\begin{equation}
T= 10^{- (A_{\mathrm{sample}} - A_{\mathrm{water}})}
\end{equation}
is normalized by the transmittance of the control sample with zero added acid ($\mathrm{ME}=0$). 
Error bars represent the standard deviation of three measurements of the same sample.
All measurements have a path length of 1.0 cm. Changing the path length shifts the onset of turbidity, but this shift can be rescaled by the path length.

\bibliography{bibtex1701} 

\end{document}